\documentclass[preprint2]{aastex}

\usepackage{natbib}
\usepackage{wasysym}
\usepackage{txfonts}
\def\AV{\mbox{A$_{\rm V}$}}

\def\nH2{\mbox{${\rm n}(\HH$)}}
\def\enH2{\mbox{$n_{(\HH$)}}}
\def\DoverH{\mbox{[D/H]}}
\def\pccc{~{\rm cm}^{-3}} 
\def\ccc{~{\rm cm}^3} 
\def\pcc {~{\rm cm}^{-2}}

\def\Tsub#1 {\mbox{$T_#1$}}
\def\TK  {\Tsub K }

\def\fH2{\mbox{f$_\HH$}}

\def\p{\mbox{$^+$}}

\def\h13cop{\mbox{{H$^{13}$CO\p}}}

\def\c3h2{\mbox{C$_3$H$_2$}}
 
 \def\R0{R$_0$}
\def\G0{\mbox{G$_0$}}

\def\ddeg{{}^\circ\kern-.1em}

\def\ps{\mbox{~s$^{-1}$}}

\def\E#1 {$10^{#1}$}
\def\E#1 {E{#1}}
\def\P#1,{$\nH2\TK~=~#1\times~10^4\pccc$~K}
\def\ec#1,#2,#3,{#1\,(#2)\E{#3}}

\def\H3{\mbox{H$_3$}}

\def\zetaH{\mbox{$\zeta_H$}}
\def\RH2{\mbox{R$_{\rm G}$}}
\def\GH2{\mbox{$\Gamma_{\HH}$}}
\def\g13{\mbox{g$_{13}$}}

\def\kHeH2{\mbox{$k_{ He-\HH}$}}

\def\tim#1,#2{\mbox{{$#1\times10^{#2}$}}}

\newcommand{\emm}[1]{\ensuremath{#1}}   
\newcommand{\emr}[1]{\emm{\mathrm{#1}}} 

\newcommand{\HH}{\emr{{\rm H}_2}}

\newcommand{\fHH}{\emr{f_{H_2}}}

\newcommand{\Zsun}{\emm{Z_\odot}}

\shorttitle{Where does HD\ come from, anyway?}
\shortauthors{H. S. Liszt }

\begin{document}


\title{HD/H$_2$ as a probe of the roles of gas, dust, light, metallicity and cosmic rays in promoting 
the growth of molecular hydrogen in the diffuse interstellar medium}


\author{H. S. Liszt}
\affil{National Radio Astronomy Observatory \\
            520 Edgemont Road,
           Charlottesville, VA,
           22903-2475}

\email{hliszt@nrao.edu}



\begin{abstract}

We modelled recent observations of UV absorption of HD and \HH\
in the Milky Way and toward damped/sub-damped Lyman alpha systems
at z=0.18 and z $>$ 1.7.  N(HD)/N(\HH)
ratios reflect the separate self-shieldings of HD and \HH\ and
the coupling introduced by deuteration chemistry.
Locally, observations are explained by diffuse molecular gas
with $ 16 \pccc \la$ n(H) $\la 128 \pccc $ if the cosmic-ray ionization
rate per H-nucleus \zetaH $= 2\times 10^{-16}\ps$ as
inferred from \H3\p\ and OH\p.  The  dominant
influence on N(HD)/N(\HH) is the cosmic-ray ionization rate
with a much weaker downward dependence on n(H) at Solar
metallicity, but dust-extinction can drive N(HD) higher as
with N(\HH).  At z $>$ 1.7,  N(HD) is comparable to the Galaxy but
with 10x smaller N(\HH) and somewhat smaller N(\HH)/N(H I).
Comparison of our Galaxy and the Magellanic Clouds shows that smaller
\HH/H is expected at sub-Solar metallicity and we show by modelling
that HD/\HH\ increases with density at low metallicity, opposite to the
Milky Way.  Observations of HD would be explained with higher
n(H) at low metallicity but high-z systems have high HD/\HH\
at metallicity 0.04 $\la$ Z $\la$ 2 Solar.
In parallel we trace dust-extinction and self-shielding effects.
The abrupt \HH\ transition to \HH/H $\approx$ 1-10\% occurs mostly from
self-shielding although it is assisted by extinction for n(H) $\la 16 \pccc$.
Interior \HH\ fractions are substantially increased by dust
extinction below $\la 32\pccc$. At smaller n(H), \zetaH,
small increases in \HH\ triggered by dust extinction can trigger
abrupt increases in N(HD).
\end{abstract}



\section{Introduction}

\def\SKL{Schmidt-Kennicutt law}

Like  molecular hydrogen \HH, the much rarer deuterated isotopologue HD has been 
studied and observed across cosmic time.  
A survey of HD/\HH\ ratios along 41 galactic sightlines has recently been published 
by \cite{SnoRos+08}, revising and greatly extending the work of 
\cite{LacAnd+05} and yet-earlier results summarized by \cite{Lis03}.  \cite{OliSem+14}
recently detected HD in a low-redshift, low-metallicity Damped Lyman Alpha (DLA) system 
with column densities N(HD) and N(\HH) very much like those seen in the Milky Way.  
HD and \HH\ have also been detected at eight redshifts toward six DLA and sub-DLA systems 
 at z $>$ 1.7  with N(HD)/N(\HH) ratios well above those seen in the Milky Way
\citep{NotPet+08,BalIva+10,IvaPet+10,NotPet+10,TumMal+10}. 

The chemistry of deuterium and HD plays a 
special role in the formation of structure and the first stars in the early Universe 
\citep{GaySta+11}.  Although it is generally understood now that observations of HD can
not provide a direct determination of the elemental [D/H] ratio 
\citep{LePRou+02,SnoRos+08,IvaPet+10}, [D/H] is well-determined by other means, 
with [D/H] $= 2.54 \times 10^{-5}$ in primordial gas
\citep{PetCoo+12,CooPet+14} and [D/H] $= 2.35 \times 10^{-5}$ locally \citep{LinDra+06}.

Given that the intrinsic [D/H] ratio is reflected only indirectly in the HD/\HH\ ratio,
the study of HD is now of interest owing to its value as a probe of the microphysics 
of the diffuse atomic gas and the more general problem of \HH-formation in relatively 
low-density diffuse neutral atomic gas.  The formation rate of HD depends primarily
on the local proton (or deuteron) density and  hence on the strength of penetrating 
hydrogen-ionizing radiation (usually the cosmic-ray ionization rate).  In turn, the 
proton density is balanced by the processes whereby  atomic ions recombine,
 basically via grain-assisted recombination mediated by the same small particles 
that heat the gas via the photoelectric effect \citep{DraSut87,BakTie94,WolHol+95}.
The interaction of H\p\ and D\p\ with HD and \HH\ to equilibrate the HD/\HH\ ratio
couples the microphysics and HD chemistry  to the general \HH\ formation problem, 
highlighting the separate roles of shielding of \HH\ and HD by themselves and by dust 
extinction.

Here we discuss these observations of HD in the context of models of the coupled 
heating/cooling- \HH\-HD- formation in diffuse neutral atomic gas.  Section 2 discusses 
models of \HH\ and HD formation and self-shielding in diffuse clouds.  In Section 3 
observations of HD in the Milky Way
are discussed and compared with the model results, which are explored in some detail 
in Section 4 in order to separate the various physical and chemical processes involved. 
Section 5 discusses what is known observationally of the H I-\HH\ transition
\footnote{In this work we refer to neutral atomic hydrogen as H I following the 
usage of \cite{SavDra+77}} in diffuse gas in nearby systems having sub-Solar metallicity and 
Section 6 discusses the observations of \HH\ and HD in high-redshift Damped Lyman Alpha 
Systems (DLA).  Section 7 is a summary.

\begin{figure*}
 \includegraphics[height=13cm]{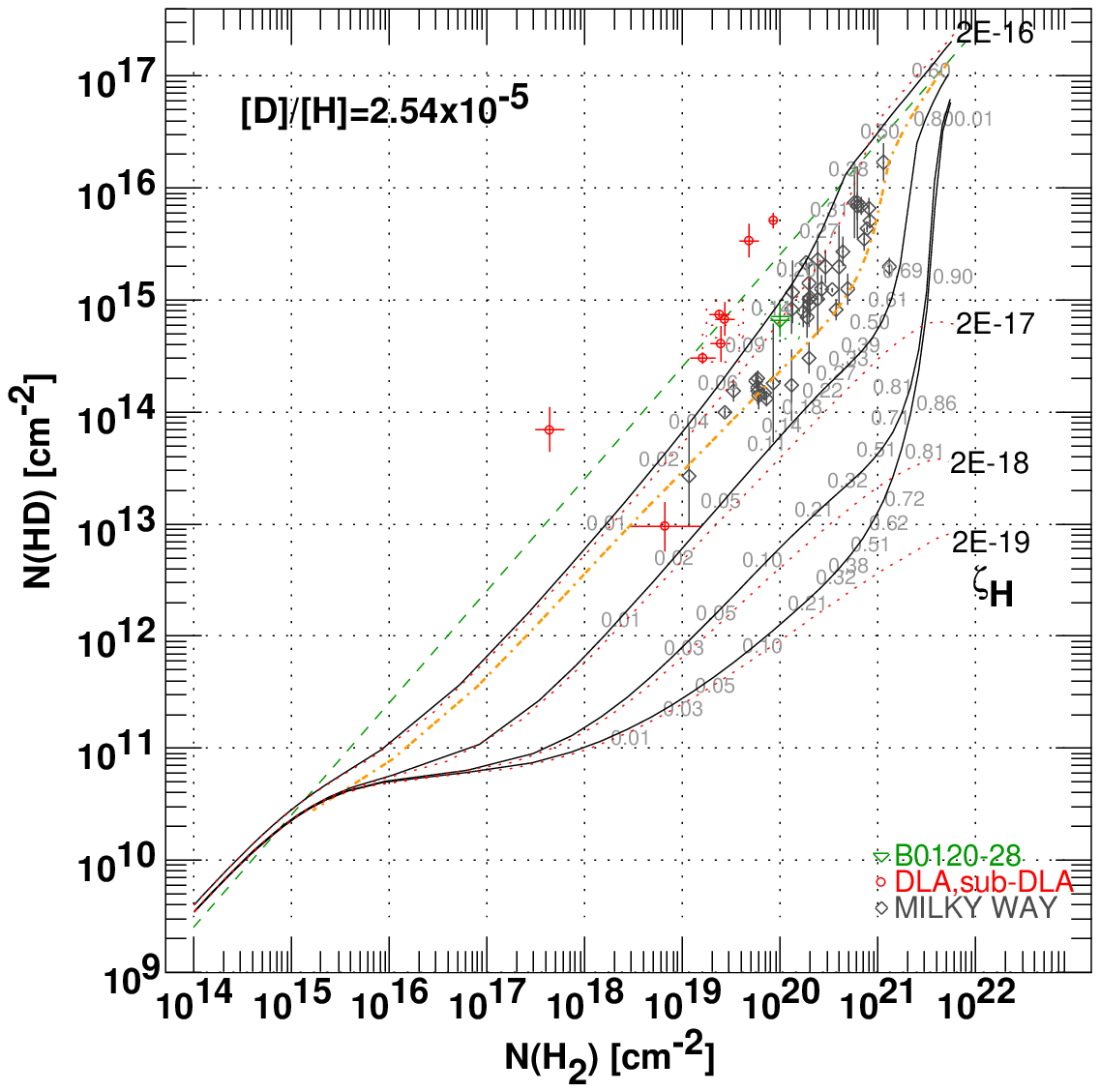}
  
\caption[]{
 Observed and model HD and \HH\ column densities for Milky Way sightlines
  from \cite{SnoRos+08} and Damped and sub-Damped Lyman Alpha systems (DLA) at 
  z = 0.18 (B0120-28) and $>$ 1.7. The solid black lines are calculated through 
 the central line of sight toward uniform density gas spheres of total density 
 n(H) = 16 $\pccc$ and primary cosmic ray ionization rates per H-atom 
 $2\times 10^{-19}\ps \le \zetaH \le 2\times 10^{-16}\ps$.  Numerical
 annotations show the \HH\ fraction along the curves. The companion 
 dotted lines show the results when the explicit shielding by dust extinction is 
 neglected.  The orange dashed-dotted line shows results for n(H) $= 128 \pccc$
 and $\zetaH = 2\times 10^{-16} \ps$ including dust extinction.  
 The primordial ratio [D/H] $= 2.54\times 10^{-5}$  \citep{PetCoo+12,CooPet+14}
 is shown as a green dashed line.
}
\end{figure*}

\begin{figure}
 \includegraphics[height=8.6cm]{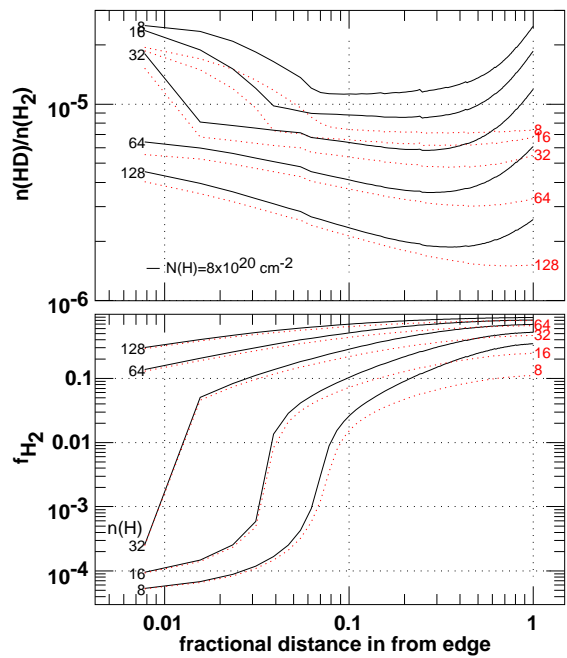}
  \caption[]{
Radial variation of the \HH-fraction (lower panel) and HD/\HH\ ratio 
(upper panel) for  models with N(H) $= 8 \times 10^{20} \pcc$ and number
density n(H) = 8, 16, 32, 64 and 128 $\pccc$.  Dotted (red) lines 
are results without dust extinction of dissociating photons.}
\end{figure}

\section{Model calculations}

\subsection{\HH\ and HD formation and self-shielding}

This work is an update of our earlier investigation of the formation of HD and 
H$_3$\p \citep{Lis03}, revised to study the wealth of new observations of HD 
noted in the Introduction.  As before \citep{Lis03,Lis07} we model the formation 
of \HH\ self-consistently in a spherical gas cloud of uniform density
immersed in the average ambient, isotropic  galactic radiation field, and we compute
the local kinetic temperature \TK\ following the methods of Wolfire and his 
collaborators \citep{WolHol+95,WolMcK+03}.  The equations of chemical and thermal 
balance are solved iteratively over a model with 128 or more equi-spaced radial 
shells, computing the radiation field in each shell averaged over the surrounding 
4$\pi$ solid angle.  

Following the prescription of \cite{Spi78} (see also \cite{SteLeP+14} 
the rate constant for \HH-formation
on grain surfaces is taken as \RH2\ $= 3\times 10^{-18}\ccc\ps \sqrt{\TK}$
but the thermal balance and temperature-dependent rate constant are not of crucial 
importance to the \HH-fraction.  The same results are obtained 
using a fixed rate constant \RH2\ $= 3.9\times 10^{-17}\ccc\ps$ that is the
average of the values obtained toward three stars by \cite{GryBou+02} 
 often cited in other work, as discussed in Section 5.2 (those sightlines 
are called out in Figure 5).  The thermal balance is however very important 
to several endothermic reactions 
driving the oxygen and deuterium chemistry and to the overall ionization
balance in the models.

The models described here differ from our previously published results 
in that they employ the \HH\ photodissociation scheme of \cite{DraBer96}
which explicitly treats dust attenuation of the radiation field at 
the wavelengths of the Lyman and Werner bands of \HH\ (90 - 110 nm).
The optical depth for dust absorption is $\tau_d$ = $1.9\times 10^{-21}$N(H)
\citep{Dra03a} as in \cite{SteLeP+14}.  Regarding our previous models based 
on the shielding factors  of \cite{LeeHer+96},
we note that incorporation of dust extinction is implicit and somewhat ambiguous
in the formulation of \cite{LeeHer+96} where only an overall \HH\
self-shielding function is employed, more similar to the earlier work
of \cite{FedGla+79}. 

The accuracy of the \cite{DraBer96} formulation  was recently been verified in 
great detail by \cite{SteLeP+14} using an exact calculation in the context
of the Meudon PDR code. Separation of dust 
extinction-related phenonema is important for understanding the HD formation 
problem, and perhaps even more important for understanding the general 
\HH-formation problem in media having low number density and/or low 
metallicity.  The separate effects of dust extinction, and the 
shielding factors and dust extinctions for our models are explicitly
shown and discussed here. 


Direct HD formation on grains is modelled following the precepts of \cite{LePRou+02}
with a rate constant 40\% larger than for \HH.  This is manifest in the
models shown in Figure 1 at very small N(\HH) but the direct formation of HD 
on grains is of almost no relevance to the observations of HD, as noted in Section 
3.  Self-shielding of HD is important in some cases, and may have contributed to
the observed HD; the self-shielding of HD is the same function of N(HD) as 
the self-shielding of \HH\  employing N(\HH) .

For reference, note that  the cosmic-ray ionization rate of atomic 
hydrogen has been taken as $\zetaH = 2\times 10^{-16}\ps$ as seems appropriate
for the diffuse molecular ISM \citep{McCHin+02,Lis03,HolKau+12,IndNeu+12}.  
As a standard value we take $\GH2\ = 4.3\times 10^{-11}\ps$ 
as the free-space photodissociation rate of \HH, as in the work of 
\cite{DraBer96} from which our \HH\ self-shielding scheme was drawn.
 
Other values in the recent literature are $\GH2\ = 2.5 \times 10^{-11}\ps$
(over 2$\pi$ steradians) in the work of \cite{LeeHer+96} and 
$\GH2\ = 5.8 \times 10^{-11}\ps$ in \cite{SteLeP+14}.
In this work, number and column densities implicitly refer to hydrogen 
nuclei when stated in the text, unless otherwise noted.  


\subsection{HD chemistry and the proton density}

HD does not enjoy the high degree of self-shielding which is the 
{\it sine qua non} of high $\HH$-fractions.  Hence the HD/\HH\ ratio  might 
be expected to be very small in diffuse gas, well below the inherent [D/H] 
ratio. That this is not so reflects the fractionation and charge exchange 
processes determined by the ambient proton and deuteron density as noted
by \cite{BlaDal73}, \cite{Wat73}, \cite{Jur74}, \cite{ODoWat74} and
\cite{Spi78}.

The basic chemistry of \HH-HD interconversion has been sketched out by those 
authors and by \cite{StaLep+98} in the context of early-Universe chemistry; 
the rates used in this work were taken from Table 1 of \cite{StaLep+98}.  Although
the fractionation/deuteration chemistry can be quite complex in cold, 
fully-molecularized dark clouds, it is fairly simple in warmer, lower-density 
diffuse regions where protons are abundant and only \HH\ and HD need be considered,
with exchange of protons or deuterons as the means of interconversion between
the two molecular hydrogen isotopologues.

In a purely atomic gas ionized by cosmic rays the ionization and 
recombination rates of H and D atoms would be very nearly the same
(the grain neutralization of D\p\ is slower by a factor $\sqrt{2}$
owing to the smaller thermal speed of the twice-heavier deuterium
isotope)
but a strong and slightly endothermic charge exchange with 
protons H\p $+$ D $+\Delta$E $\rightarrow$ D\p $+$ H
(rate constant $k_1 = 10^{-9}~\ccc$\ps, $\Delta$E/k = 41 K) 
tends to force n(D\p)/n(D I) $\approx$ n(H\p)/n(H I) exp(-41 K/T).  In the 
presence of $\HH$ a rapid and relatively strongly exothermic 
reaction D\p\ $ + \HH \rightarrow$ HD + H\p\ forms HD with rate
constant  $k_2 = 2.1 \times 10^{-9}~{\rm cm}^3$ \ps .

If only charge transfer and \HH\ fractionation neutralize D\p\ 
(a highly reductive assumption), a  relatively 
compact expression gives the proton density n(p) required to reproduce a 
given HD/\HH\ ratio in terms of observed quantities and physical constants, 
with no explicit dependence on either the density or recombination rates: 
we have 

$$ n(p) = {{ {{\rm n(HD)}}/{\rm n}(\HH)} \over {\rm [D/H]}}
  ~ { \Gamma_{\rm HD} \over{{\rm k}_2}} 
     [1+({{{\rm k}_2}\over {{\rm k}_1}}-2) {{{\rm n}(\HH)}\over{\rm n(H)}}] 
    \exp{ ({ {41} \over T}) } 
\eqno(1) $$

where n(H) = n(H I) + 2 n($\HH$) and the photodissociation rate of HD in free 
space is $\Gamma_{\rm HD} = \Gamma_{{\rm H}_2} = 4.3\times10^{-11}$\ps\ 
\citep{DraBer96,LePRou+02}. 
The required proton density n(p) derived from Eq. 1 is nearly independent 
of the molecular fraction in the gas for k$_2/$k$_1 = 2.1$ and there
is no explicit dependence on density if the \HH\ fraction is fixed.  
Of course the actual proton density may have quite strong dependence on n(H)
and the assumptions used to derive Eq. 1 are rather archaic.  Below
we discuss the actual proton density in the models but the chief means
by which an adequate proton density is achieved is the high
default  cosmic ray ionization rate that we have adopted, see 
\cite{Lis03}.

\subsection{The D/H ratio}

The models whose results are shown here use the cosmic ratio 
\DoverH\ $= 2.54 \times 10^{-5}$  \citep{PetCoo+12,CooPet+14}
which is near the Milky Way value \DoverH\ $= 2.35\pm 0.24 \times 10^{-5}$ 
determined by \cite{LinDra+06}.  The actual gas phase \DoverH\ may be slightly 
smaller than the overall \DoverH\ value in the Milky Way but this is a 
small  difference compared to the effects of the chemistry.  Moreover 
the model results are intended to be generally relevant, for instance in Figure 1 
where the local and high-z results are shown together.

\section{Observations and models of HD in the Milky Way}

\subsection{Observations of HD}

Shown in Figure 1 are the observational results for N(HD) and N(\HH) along 
the 41 Milky Way sightlines in the recent  omnibus FUSE survey of 
\cite{SnoRos+08}.  Results for the DLA sightlines at z $>$ 1.7 shown 
in Figure 1 are summarized in Table 1 and discussed in Section 6.  
Data for the low-z, low-metallicity DLA system at z=0.18 
discussed by \cite{OliSem+14} are also shown in Figure 1.

In a gas where H and D exist primarily as \HH\ and HD, N(HD)/N(\HH) $\approx$ 2[D/H] 
$\approx 5 \times 10^{-5}$.  In the opposite limit when only insignificant amounts
of H and D are molecular, N(HD)/N(\HH) $\approx 1.4$ [D/H] \citep{LePRou+02}.  
By contrast the galactic HD column densities lie about a factor 10 below the cosmic 
\DoverH\ ratio in Fig. 1, falling nearly 
parallel to a line of constant N(HD)/N(\HH) = $3\times 10^{-6}$.
The regression analysis of \cite{SnoRos+08} found a power-law slope $1.25\pm0.03$.
The N(HD)/N(\HH) values of \cite{SnoRos+08} are about three times larger than
those considered in our simillar analysis of the same phenomena \citep{Lis03}.

\subsection{Comparison with models}

The slightly super-linear empirical slope determined by \cite{SnoRos+08}
means that N(HD)/N(\HH) increases with increasing
molecular fraction \fH2\ = 2N(\HH)/N(H) with N(H) = N(H I)+2N(\HH) (note the
annotations in Fig. 1 showing \fH2\ along the curves).   
\cite{SnoRos+08} pointed out that the models of \cite{LePRou+02}
seemed to predict the opposite behaviour except at \fH2\ $\ga$ 0.9 and 
further noted that those models generally
underpredicted N(HD) unless \fH2\ $\la$ 0.1 or \fH2\ $\ga$ 0.9.  This
was a straightforward consequence of the gas-phase chemistry of HD that
 segregated HD in regions of smaller density and lower \HH-fraction
in diffuse clouds because of its dependence on the presence of a relatively
high proton density.

Shown in Figure 1 are our equilibrium model results for sightlines through the centers 
of the uniform-density gas spheres discussed in Section 2.  Results for a family 
of models with n(H) = 16 $\pccc$ and varying primary cosmic-ray ionization rate 
per H-atom 
$2\times 10^{-19}\ps \le \zetaH \le 2\times 10^{-16}\ps$ are shown with and 
without the attenuation of Lyman and Werner Band photodissociating radiation
by dust. Also shown  are model results for n(H) = 128 $\pccc$.

At the lowest cosmic-ray ionization rate considered, $\zetaH = 2\times 10^{-19}\ps$,
the chemistry is essentially switched off and the predicted HD abundance is some 
three orders of magnitude below observed values.  This demonstrates that when
HD is seen in the Milky Way it overwhelmingly originates {\it in situ} in the 
gas phase by deuteration of \HH.  Although the actual [D/H] ratio in the gas 
phase is important, and n(D)/n(H) in the gas phase is affected by the 
n(HD)/n(\HH) ratio \citep{Lis06DoverH}, considerations of HD formation on 
grains are irrelevant to HD-formation in the observed amounts.  HD molecules formed 
on grains comprise a negligible fraction of the observed HD.

At the highest cosmic-ray ionization rate considered, 
$\zetaH = 2\times 10^{-16}\ps$, the observations are bounded by the 
models with  n(H) = 16 $\pccc$ and 128 $\pccc$.  If the cosmic ray ionization 
rate is ubiquitous, the region  occupied by the observations in Figure 1 
is understood in  terms of the relatively slow variations of the N(HD)/N(\HH)  ratio 
with number density, and clouds  with n(H) $>>$ 128 $\pccc$ apparently were not 
sampled in the observations.  Alternatively, if the cosmic ray ionization rate 
is assumed to vary, it should generally be in the sense of increasing above 
$\zetaH = 2\times 10^{-16} \ps$, implying somewhat higher number density, 
because neutral atomic gas with number density much below $ 16 \pccc$ is 
not generally understood to be thermally stable in multi-phase models of 
the ISM , see e.g. Figure 7 of \cite{WolMcK+03}.

Our models with  $\zetaH = 2\times 10^{-16}\ps$ and 
$16\pccc \la$ n(H) $\le 128\pccc$ reproduce the observations, including the 
slightly super-linear slope derived by \cite{SnoRos+08}, but the slope of 
the observed variation of N(HD) with N(\HH) must also reflect the underlying 
distribution of physical environments that were sampled.
The HD column density results from a variety of influences including the
gas-phase chemistry, the individual self-shielding of HD and \HH,
their coupling via the chemistry, and the extinction of photodissociating 
UV radiation by dust.  The consequential difference between our models and those of 
\cite{LePRou+02} lies most nearly in the treatment of the overall ionization 
balance in the presence of grain-assisted recombination, which both allows and 
requires a higher cosmic ray ionization rate.

\subsection{Cloud structure in \fH2\ and n(HD)/n(\HH)}

Although our models have uniform total number density n(H), the molecular fraction, 
proton density and n(HD)/n(\HH) ratio vary considerably within them.  Figure
2 shows the n(HD)/n(\HH) ratio and molecular fraction as functions of radius
for models with N(H) = $8\times 10^{20}\pcc$ having column densities  N(HD)
and N(\HH) that are typical of the Milky Way data shown in Figure 1.  

The models
have 128 radial shells but only barely resolve some sharp variations near the
outer cloud edges.  Typical variations in n(HD)/n(\HH) are a factor two or three
with higher values at the edge and center.   At the highest density, n(H) = 
$128 \pccc$,  the \HH-fraction varies less and generally in the opposite sense 
from n(HD)/n(\HH) so that HD is distributed rather uniformly throughout the model.
At the lowest density, n(HD)/n(\HH) varies much less than the molecular
fraction and the bulk of the HD resides near the center of the model.

Two effects directly attributable to dust extinction are shown in Figure 2.
First, the n(HD)/n(\HH) ratio is slightly smaller in the outer regions of the
models when dust attenuation is negelected, presumably because the attenuation
by dust has a larger effect on HD than on \HH, given that HD is so much more
weakly self-shielded.  Second and most noticeably, the n(HD)/n(\HH) ratio does 
not rise near the center of models in which the extinction of dissociating photons 
by dust is neglected (but note that the scale in the upper panel of
Figure 2 is much expanded compared to that at bottom).  An analogous
situation appears in Figure 1 for smaller values of \zetaH\ when N(HD) 
only rises sharply at N(\HH) $\ga 10^{21}\pcc$ in the presence of dust 
extinction.  Although the \HH-fraction always reaches higher values 
toward the center of models when the explicit attenuation of dissociating
photons by dust is included, sharp increases in n(HD)/n(\HH) and  N(HD)/N(\HH)
result primarily from the onset of strong HD self-shielding, triggered by the
extinction due to dust.  Whether the onset of \HH\ self-shielding and high
\HH-fractions is triggered by dust extinction is the subject of Section 4.

\begin{figure}
 \includegraphics[height=6.8cm]{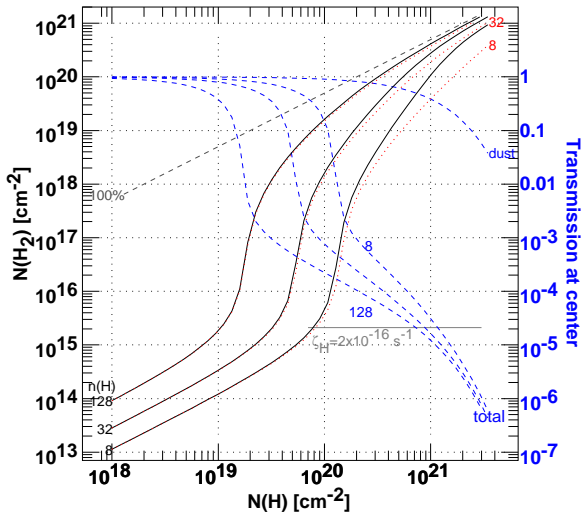}
  \caption[]{
Calculated \HH\ column densities and interior attenuations.  The
solid black curves represent the \HH\ column density toward the
geometric centers of models with n(H) = 8, 32 and 128 $\pccc$
over a wide range of cloud column density N(H) from front to
back of the model, with $\zetaH = 2\times10^{-16}\ps$. 
As in Figure 1 each solid curve has a dotted
red companion representing the same model neglecting dust 
attenuation.  Reading the scale at right,  the blue dashed
curves separately show the fraction of ambient \HH-dissociating photons 
that penetrates to the centers of the models after \HH\ self-shielding
and dust extinction (labelled ``total'') and the much smaller amount 
which the dust extinction contributed to that total. 
The grey horizontal line represents the 
attenuation at which the cosmic ray flux indicated becomes the 
dominant cause of \HH\ destruction at the rate indicated.
}

\end{figure}

\section{What triggers the onset of high \HH\ and HD fractions?}

\subsection{The outermost self-shielding layer}

 In this work, dust extinction of dissociating photons is expressed using 
a single absorption cross-section $\sigma_d = 1.9 \times 10^{-21} N(H)$ as in 
the work of \cite{DraBer96} and \cite{SteLeP+14} and scattering is ignored.
As shown in the Figure 2, dust extinction raises n(HD)/n(\HH) 
and \fH2\ near the centers of the models but does not play a large role 
in triggering the onset of higher n(\HH) in the outer shielding layers except 
perhaps at the very lowest density: the curves showing the radial variation 
of \fH2\ with and without dust extinction differ little in the outer radial 
regions where \fH2\ abruptly increases by factors of 100 or more.  The reason 
for this can be better understood in the context of  Figure 3 showing the 
attenuation of dissociating photons in models 
with differing  number and \HH\ column density, along with the particular 
contribution  from dust extinction of dissociating photons.

In Figure 3 the onset of higher \HH\ fractions is scarcely affected by the 
presence of dust attenuation, rather, that is determined by the onset of 
strong \HH\ self-shielding which already occurs at such small \AV\ that 
there is very little dust opacity even in the  UV.  If  physical conditions 
foster small \HH-fractions at high N(H), the dust extinction and its effect 
on \fHH\ may be appreciable for \HH\ and this is the situation for HD as 
well at very high N(H) in Figure 1.  The details depend on 
the slopes of the variation of the attenuation with column density:  
if the self-shielding of \HH\ varies slowly with N(\HH), especially 
when \fHH\ is larger, dust attenuation and small increases in N(H) 
may be efficient at increasing the molecular fraction locally.  At 
the lowest number density in Figure 3, N(\HH) toward the center of the 
models about doubles for N(H) $\ga 10^{21}\pcc$ and the strong
increases in HD were noted earlier. Whether clouds with 
such high N(\HH) exist at such low density is another matter.

Also shown in Figure 3 is the transmission at which the attenuation of the
radiation field is so great that the destruction rate of \HH\ by 
cosmic rays equals that due to photodissociation in the Lyman and 
Werner bands when $\zetaH = 2\times10^{-16}\ps$. In fact
the destruction of \HH\ near the centers of our models is dominated by 
cosmic rays even for clouds having total \AV\ = 0.5 - 1 mag (see
also Section 5.2).
The high cosmic-ray ionization rates required to explain HD strongly
limit the ability of dust extinction to increase \fH2\ inside the
outer self-shielding layer.

\subsection{The overall contribution from dust extinction}

The overall contribution of dust extinction of Lyman and Werner band photons
is summarized in Figure 4 where for models with n(H) = 8, 16, ..., 128 $\pccc$
we show the total molecular fraction integrated over the model and the fraction 
of that total that is directly attributable to dust extinction, calculated 
symbolically as ((Mass with dust extinction)-(Mass without))/Mass(with).
For conditions approximating a Spitzer Standard H I cloud with 
n(H) = 32$\pccc$ and N(H)$ = 4\times 10^{20}\pcc$ some 20\% of the hydrogen
is in \HH\ and 30\% of that is attributable to the presence of dust extinction.

\section{\HH\ formation at Solar and sub-Solar metallicity}

\subsection{Observations}

In Figure 5 we show N(\HH) and N(H) as determined in UV absorption in
the Milky Way and the Magellanic Clouds.  The galactic data include
observations toward
bright stars as studied by Copernicus \citep{SavDra+77} and toward bright
stars \citep{RacSno+02,RacSno+09} and AGN \citep{GilShu06} using FUSE, extending
to much higher N(H) than were available to Copernicus.  The Southern
Hemisphere data is from \cite{TumShu+02}. 
The Milky Way data in this plot shows the familiar jump in N(\HH) at 
N(H) $\approx 2-5 \times 10^{20}\pcc$ whereby the fraction of
hydrogen in \HH\ abruptly increases to $\ge 0.01$.  

Plotting N(\HH) against N(H) is unlike the situation shown in Figure 1, where
sightlines have been ordered according to their \HH\ column densities.  
Viewing the onset of \HH\ formation along sightlines harboring 
a mixture of conditions involves several separate aspects 
of the ISM; the cold neutral medium of the ISM is clumped into diffuse 
``clouds'' (usually called H I clouds by radio astronomers);  \HH\ forms 
in appreciable quantities in H I clouds having sufficiently high N(H); and 
lines of sight with N(H) $\ga 3\times 10^{20}\pcc$ cross at least one of 
these clouds.  The  column densities at which the jump occurs 
has been increased, and somewhat spread out by the contribution from 
unrelated, less-molecular material along the line of sight, see 
\cite{Spi85}.

The sightlines used  by \cite{GryBou+02} to determine \RH2\ are 
marked in Figure 5; they all have very high \HH-fractions at their respective 
values of N(H): nonetheless the value of the grain surface \HH-formation
rate constant derived in that work reproduces our temperature-dependent
results extremely well as noted in Section 5.2 and as shown in Figure 6.

For the Large Magellanic Cloud (LMC), the metallicity is some 2.5 times smaller 
than that of the Milky Way (Z=0.007 vs. 0.018 \citep{Duf84}) and the jump in 
\HH\ occurs around  N(H) = $10^{21}\pcc$, roughly three times that of the Milky 
Way.  According to \cite{WeiDra01Z} such N(H) corresponds to \AV $\approx 0.12$ mag, 
as against \AV\ = 0.16 mag in the Milky Way.  For the Small Magellanic Cloud (SMC) 
with Z = 0.002  \citep{Duf84}, the jump in \HH\ occurs at even higher 
N(H) $\approx 3\times 10^{21}\pcc$, but corresponding
to \AV\ = 0.18 mag \citep{WeiDra01Z}, again very similar to the Milky Way.

\begin{table*}
\caption[]{HD in high-z DLA and sub-DLA systems}
{
\begin{tabular}{lcccccccc}
\hline
Source& Z/\Zsun    & z & N(H I)        & N(\HH)     & N(HD)      & 2N(\HH)/N(H I) & N(HD)/N(\HH) & HD Ref. \\
      &            &   & log $\pcc$  & log $\pcc$ & log $\pcc$ &      &   /[D/H] & \\   
\hline
Q1232   & 0.04     & 2.34       & 20.90(0.08) & 19.68(0.08) & 15.43(0.15) & 0.121 & 2.21 & 1\\
Q1331a  & 0.04     & 1.78       & 21.20(0.04)$^b$ & 19.43(0.10) & 14.83(0.15) & 0.034 & 0.99 & 2\\
Q1331b  &          &            &             & 19.39(0.11) & 14.61(0.20) & 0.031 & 0.65 & 2 \\
FJ0812a & $\la$0.36$^a$ & 2.63       & 21.35(0.10) & 19.93(0.04) & 15.71(0.07) & 0.076 & 2.37 & 2 \\
FJ0812b &          &            & 21.35(0.10) & 18.82(0.37) & 12.98(0.22) & 0.006 & 0.06 & 2 \\
J2123   & 0.5      & 2.06       & 19.18(0.15) & 17.64(0.15) & 13.84(0.20) & 0.058 & 6.20  & 3\\ 
J1439   &  1       & 2.42       & 20.10(0.10) & 19.38(0.10) & 14.87(0.03) & 0.380 & 1.22 & 4\\
J1237   & 2        & 2.69       & 20.00(0.15) & 19.21(0.13) & 14.48(0.05) & 0.324 & 0.73 & 5\\
\hline
\end{tabular}}
\\
$^a$ \cite{ProHow+03,BalIva+10}\\
$^b$ \cite{ProWol99} \\
References (1) \cite{IvaPet+10}; (2) \cite{BalIva+10}; (3) \cite{TumMal+10}; (4) \cite{NotPet+08}; (5) \cite{NotPet+10} \\
\end{table*}

\subsection{Modelling the onset of \HH\ self-shielding}

In Section 5.1 we showed that the surface area of large grains (as represented 
by \AV\ in the small range  0.12 mag $\la \AV \la 0.18$ mag) is very nearly 
constant at the onset of strong self-shielding in
\HH\ in three systems of different metallicity, in principle leading to 
the question whether it is the extinction by these grains or their aggregate 
\HH-catalytic grain surface area that is responsible for the increase 
in the \HH-fraction with N(H).  


\begin{figure}
 \includegraphics[height=7.6cm]{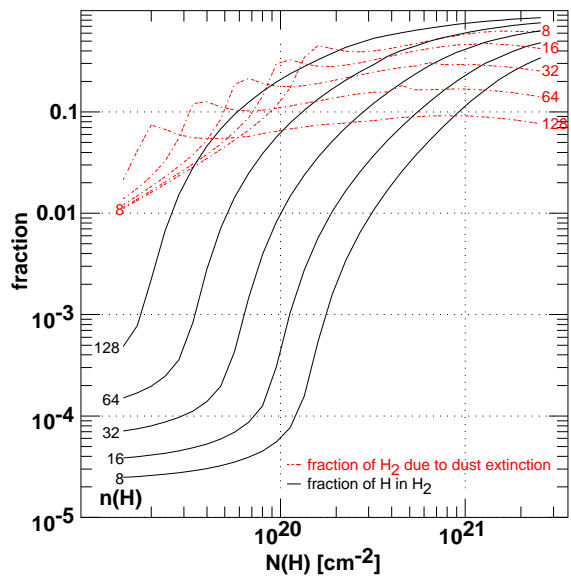}

  \caption[]{Molecular mass fractions and fractions of the molecular
  mass attributable to dust extinction.  The solid curves show the 
 fraction of the cloud model mass that is molecular at the indicated
  number and column density.  The dash-dotted red curves nearer
the top of the plot show
the fraction of the molecular mass that is attributable to dust 
 extinction of Lyman and Werner Band photons as  discussed in Section 
 4. } 
\end{figure}

The behaviour seen in Figure 3 suggests that only the grain
catalytic area is important, because the dust extinction 
{\it per se} is small at the onset of  \HH\ self-shielding.
This is further illustrated in Figure 6 showing the results of 
varying several parameters in models with n(H) $= 32\pccc$.  
This is the density of a Spitzer H I cloud and
for typical ISM pressures p/k = $2-3000 \pccc$ K \citep{JenTri11}
it is consistent with the kinetic temperatures that have been
inferred from the J=1 and J=0 levels of \HH, ie 77 K for
the original Copernicus survey \citep{SavDra+77}, $86\pm20$K and
$124\pm8$K for FUSE sightlines toward distant AGN in the galactic disk 
and halo, respectively \citep{GilShu06} and $67\pm15$K for the
highest column density translucent FUSE sightlines toward bright stars
 \citep{RacSno+02,RacSno+09}.  The kinetic temperatures in our models
are in the range 50 - 160 K, varying inversely with n(H) and having 
somewhat higher pressure at higher density, as is generally the
case for phase diagrams in multi-phase models of the
diffuse ISM.  This is all consistent with the heating-cooling model that 
we adopted and is discussed in greater detail by \cite{WolMcK+03}.

The curves shown in Figure 6  cluster in two groups according to
whether the H-\HH\ transition is shifted appreciably. Some
parameters have little effect. Replacing
our temperature-dependent \HH\ formation rate by the value derived
by \cite{GryBou+02} has little effect.  Increasing the cosmic-ray
ionization rate to $\zetaH = 10^{-15}\ps$ increases \HH\ formation 
and hastens the onset of \HH\ self-shielding at small N(H) because 
the models are somewhat warmer but \HH\ formation is suppressed at the very
highest N(H) because cosmic ray ionization so greatly dominates
the \HH\ destruction.  The column density at the onset of \HH\ self-shielding
scales inversely with the radiation field as expected from the discussion
in Appendix B but dependence on the radiation field is extremely complex
because the thermal and ionization balance is strongly affected.  Stronger
radiation heats the models, increasing the \HH\ formation rate, somewhat
compensating the increased photodissociation.

Also shown in Figure 6 are curves corresponding to decreasing the grain
surface \HH-formation rate constant \RH2  by a factor of 10, and separately, 
a calculation depleting the quantity of large, \HH-forming grains by the same 
amount and perforce also lessening the extinction by dust of dissociating 
photons by the same amount.  Figure 6 shows that the onset of \HH\ 
self-shielding in the outer shielding layer depends linearly on the 
grain surface area, but only to the extent that this area is available 
for catalytic \HH\ formation; reducing the grain formation rate and the 
surface area of large grains both result in self-shielding at very 
nearly the same N(H), confirming that the 
extinction provided by the grain surface area is a small effect on
the initial onset of strong \HH\ self-shielding and the H I - \HH\ transition.

The effects directly attributable to the grain surface area differ
noticeably inside the self-shielding layer.  When large grains
are removed entirely the \HH-fraction does not exceed 10\% except
at very large N(H), which is the case in Figure 5 for the sightlines
with sub-Solar metallicity.  Inside  the self-shielding layer, dust 
extinction drives the \HH-fraction higher, and dust extinction will 
have a strong effect whenever the \HH\ fraction is substantially 
below unity well inside the 
outer self-shielding layer; this would be the case at low n(H) and large N(H) if 
those conditions actually occur in the ISM.  This has interesting 
consequences for the time-evolution of the \HH\ fraction, because 
regions with lower molecular fraction equilibrate earlier, 
and, even though  dust extinction can increase the equilibrium
molecular fraction inside the self-shielding layer, it does not hasten 
the approach to equilibrium.  When dust extinction matters to the
\HH\ fraction  the times to reach \HH\ equilibrium will be  long.  

\begin{figure*}
 \includegraphics[height=13cm]{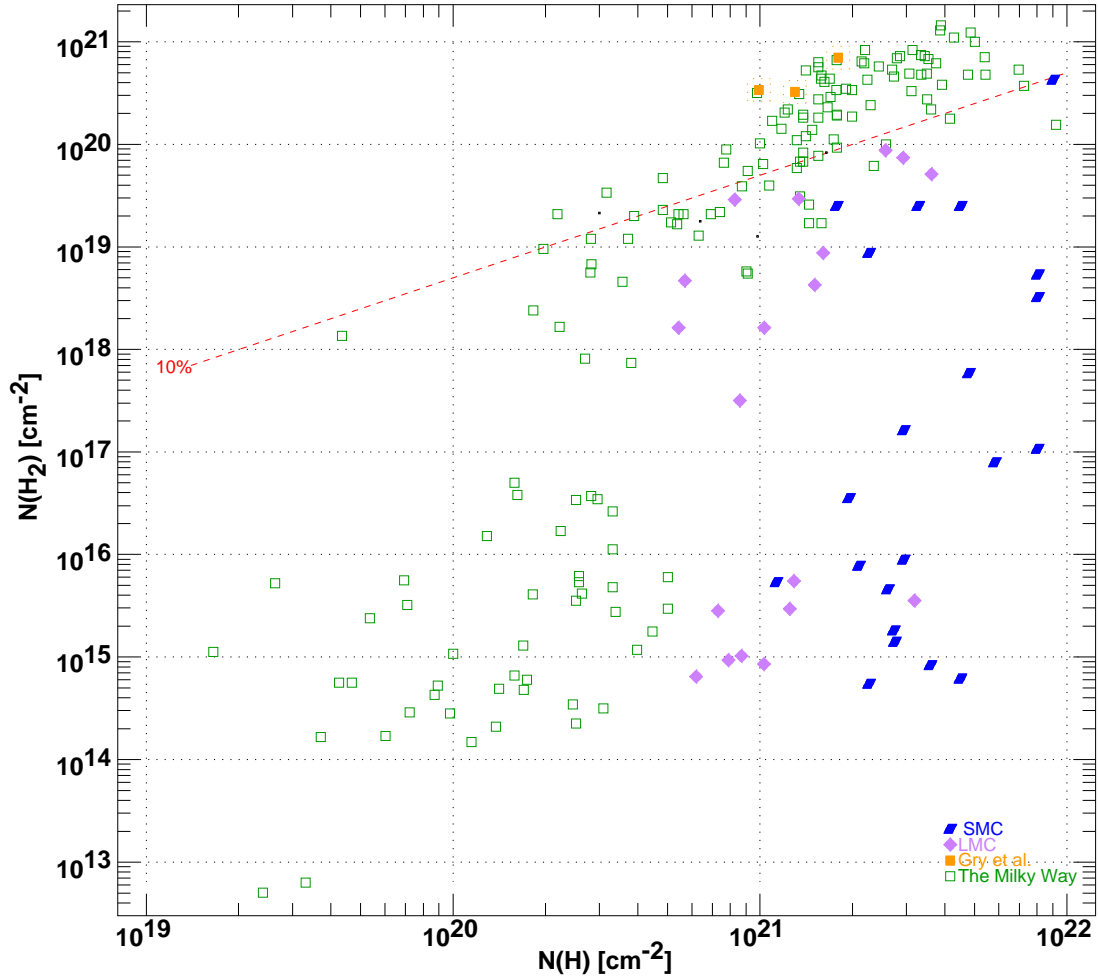}
  
\caption[]{
 \HH\ column densities for the Milky Way and Magellanic Clouds.  The Milky
  Way data include results from \cite{SavDra+77}, \cite{RacSno+02}
 and \cite{GilShu06} and are shown as open green rectangles;  the 
 three lines of sight used by \cite{GryBou+02}
  to determine the \HH-formation rate coefficient are separately
  marked.  Results for the SMC and LMC are taken from \cite{TumShu+02}.
}
\end{figure*}

\begin{figure}
 \includegraphics[height=7.13cm]{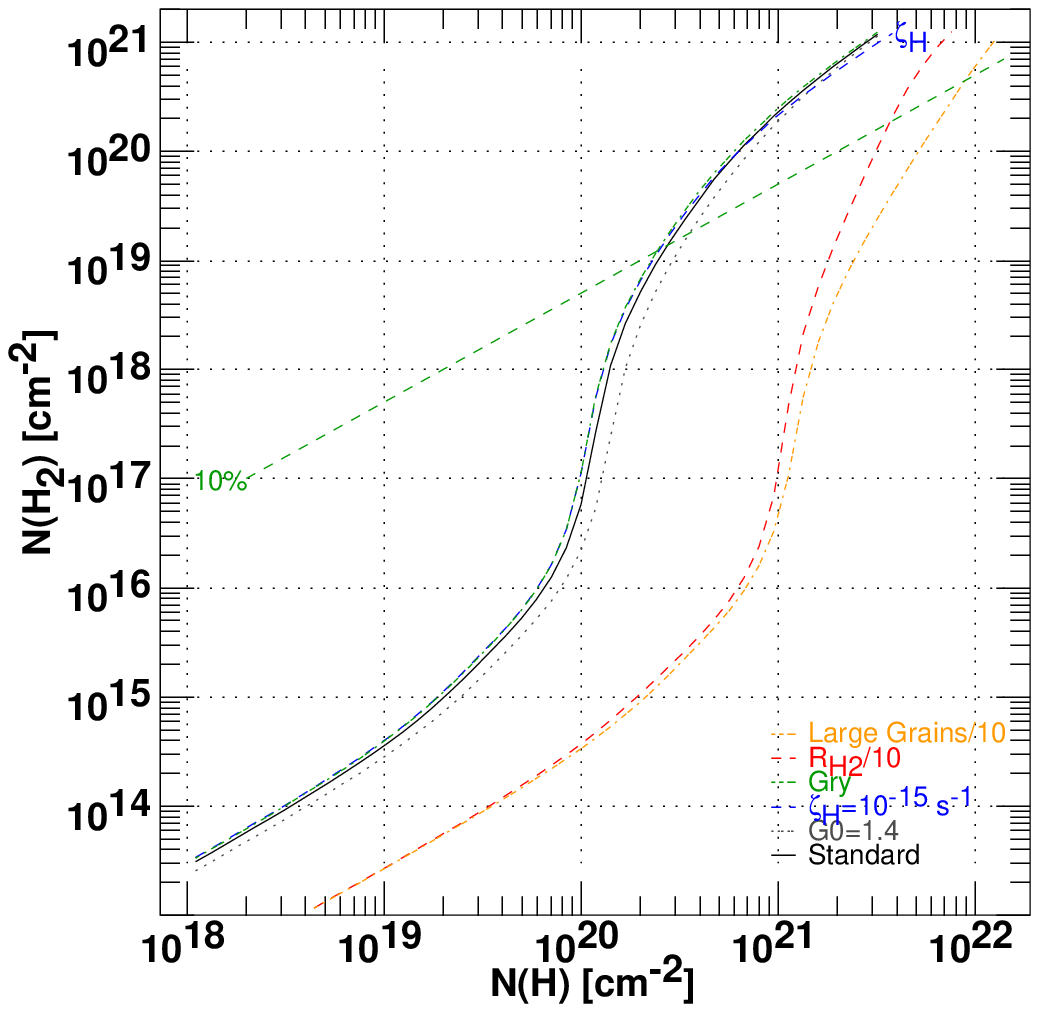}
\caption[]{Other influences on N(\HH) for models at n(H) = $32\pccc$.  The 
group of curves at the left includes  a ``standard'' model (solid black), 
a model with the radiation field increased by 40\% (dotted grey), a model 
using the temperature-independent \HH-formation rate of \cite{GryBou+02}
(dash-dot green) and a model with $\zeta_{\HH} = 10^{-15}\ps$ (dashed, blue).  
At right are models with a diminished grain formation rate coefficient 
(dashed red) and a model in which 90\% of the large grains are entirely 
removed from the calculation (dash-dot orange).  The green dashed line 
indicates the locus where 10\% of the hydrogen is in molecular form.
These models were computed along sightlines displaced 5/6 of
the radius from the center, compare with Figure 3 to see
that the H I - \HH\ transition has been displaced toward 
higher N(H) compared with models observed through the central
sightline.}
\end{figure}

\section{HD formation at high redhshift and/or low metallicity}

The observations of HD in DLA and sub-DLA systems at high redshift are summarized 
in Table 1 and shown in Figures 1 and 7.  Although DLA and sub-DLA systems typically have
low metallicity, the systems in which HD has been detected cover a wide range of
metallicity from Z/\Zsun = 0.04 to 2.  It is relatively easy to explain the high
N(HD)/N(\HH) seen at high z in terms of higher density when the metallicity is
small and rather puzzling that the high redshift systems are so similar in 
having such high  HD/\HH\ ratios while the metallicity varies over such a wide range. 

Figure 6 shows that a given N(\HH) will be seen at smaller molecular fractions 
in systems of lower metallicity because the required N(H) is larger; indeed,
the DLA and sub-DLA systems at high redshift have comparable N(HD) to those seen 
in the Milky Way sightlines, but generally at smaller molecular fractions overall. 
In Table 1 systems with high and low metallicity have comparable values of N(\HH) 
but with much smaller \fH2\ at the low-metallicity end.
 
Unlike the case at Solar metallicity illustrated in Figure 1 where N(HD)/N(\HH) 
decreases with density, larger proton densities and higher HD/\HH\ ratios are 
expected at {\it higher} number density when the metallicity is small.  Higher proton 
densities are always more readily available in principle at higher number
densities but radiative recombination with electrons 
from ionized carbon and neutralizations by small grains suppress the proton 
fraction at higher density  when the metallicity is near-Solar.  When the metallicity 
is small, the effects of both electrons and small grains are sharply 
curtailed and the proton density better tracks the number density overall.

The situation is illustrated in Figure 7 
where we varied the metallicity in our  calculations by linearly scaling 
the abundance of all metals or metal-bearing species (ie the large and small
dust grains).  At left it is shown that the N(HD)/N(\HH)
ratio increases at least as fast as 1/Z at n(H) $ = 128 \pccc$ in the range of N(\HH) 
where HD is observed at high redshift.  At right in Figure 7 the increase of
N(HD)/N(\HH) with density is shown for Z/\Zsun = 1/16.  The functional dependence
on density is weaker than linear but  the senseis quite contrary to
the variation with density shown in Figure 1 at Solar metallicity.

\begin{figure*}
 \includegraphics[height=7.13cm]{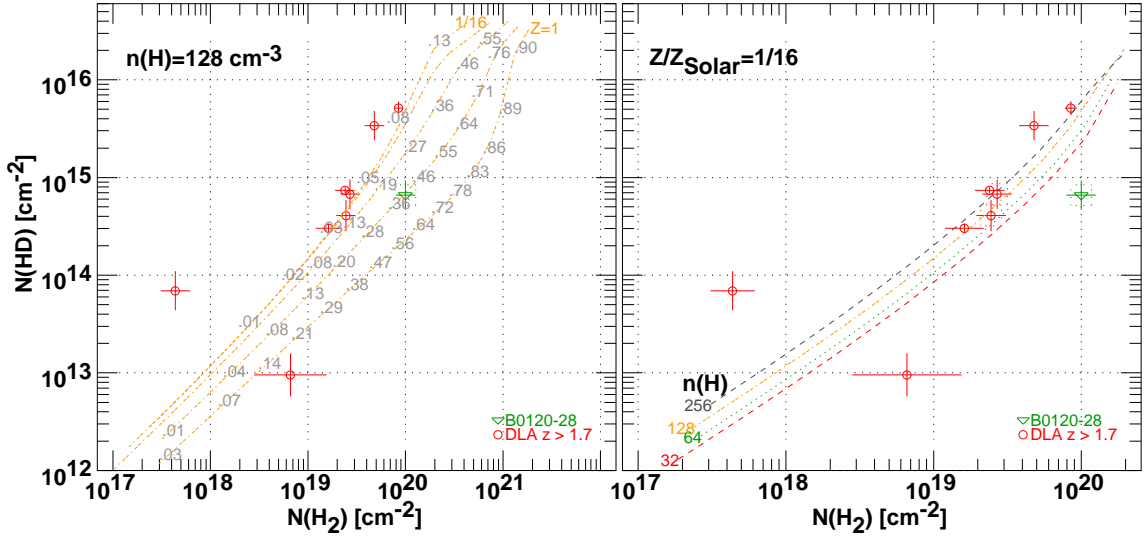}
\caption[]{Influence of density and metallicity on N(HD)/N(\HH). 
Left: at n(H) $= 128\pccc$ the metallicity is varied in steps of two
from Solar to 1/16 Solar.  Values of \fH2\ are indicated in gray along
the curves.  Right:n(H) is varied  in steps of two over the range
32 $\pccc \le $ n(H) $\le 256 \pccc$ at 1/16 Solar  metallicity. }
\end{figure*}

\subsection{n(H) inferred for J0812+3208A}

\cite{BalIva+10} derived \TK\ $= 48 \pm 2$ K from comparison of the J=1 and J=0 
levels of \HH\ toward J0812+3208A and observed a ratio log N$_1$/N$_0$ = -1.93 
in the J=1 and J=0 levels of HD.  Unlike \HH\ the J=1-0 transition of HD is
permitted, with a small dipole moment and a spontaneous transition rate
A$_{10} = 5.1 \times 10^{-8} \ps$ \citep{FloLeB+00}.  \cite{BalIva+10} 
pointed out that the HD level populations can be used to derive the 
ambient density and they quote n  $\approx 50 \pccc$ using a two-level atom 
approach under the assumption that radiative pumping of the J=1 level is 
negligible.  They made several errors in their discussion, most importantly 
using the downward collision rate C$_{10}$ in their unnumbered expression 
for the number density (their  Section 3.2) when the smaller upward rate is actually 
called for.

The collisional rate constants for HD excitation by H, He, and \HH\ are
given by \cite{FloLeB+00}.  They allow a number of important simplifications
in the analysis below 100 K, because the rate constants are the same for H and He, 
and the same for ortho- and para-\HH.  As well the rate for \HH\ excitation
is just twice that for atomic hydrogen so that the analysis is independent of the 
molecular fraction.  Denoting the upward rate constant for excitation by atomic hydrogen
as $C_{01}$ and the spontaneous emission coefficient as A$_{10}$, the total hydrogen 
number density may be written as

$$ n(H) = {N_1 \over N_0} {A_{10}\over C_{01}} \times {1\over{1+[{\rm He}/{\rm H}]}} $$

$$ C_{01} = {g_1 \over g_0} e^{-E_{10}/{kT_K}} C_{10} = 3 e^{-128/T_K} C_{10} $$

$$ C_{10} = 3.2\times 10^{-12} {\rm cm}^3 \ps T_K^{1/3} $$

in the limit of weak excitation n(H)$C_{01} << A_{10}$.

With \TK\ = 48K and [He]/[H] = 0.085 by number, we find n(H) = 240 $\pccc$.  Such a density 
is consistent with observed, relatively high HD/\HH\ at low metallicity (Figure 7)
and indeed \cite{BalIva+10} considered that log Z/\Zsun = -1.
However \cite{ProHow+03} actually derived [O]/[H] $\approx $ 0.36 Solar and similar
results for [Zn]/[H], so that J0812 is {\it not} an especially low-metallicity system.

\section{Summary}

\subsection{HD}

In Section 3 (see Figure 1) we showed that the ratios N(HD)/N(\HH) $\approx 3\times 10^{-6}$ 
in the Milky Way are primarily diagnostic of the cosmic-ray ionization rate and 
secondarily of the number density in the sense of \zetaH/n(H) tending to be constant, 
with $16 \pccc \la $ n(H) $\la 128 \pccc $ for \zetaH\ = $2\times 10^{-16}\ps$.  
The gas sampled in galactic HD measurements has moderate density and molecular 
fraction \fH2\ $\approx 0.1 - 0.5$, because the proton densities necessary 
to ensure adequate protonation are more difficult to maintain at the higher densities
that are more favorable to \HH\ formation overall (see Figures 2 and 8).

In Section {\ bf 6} we discussed N(HD) and N(\HH) observed in high-redshift DLA and sub-DLA 
systems. DLA and sub-DLA systems observed at z $>$ 1.7 have much higher N(HD)/N(\HH) 
ratios compared to the Milky Way, ie with comparable N(HD) but order of magnitude smaller 
N(\HH) and somewhat smaller N(\HH)/N(H I).  In Section 5 (Figure 5) we discussed how 
smaller \HH\ fractions are observed in nearby systems with sub-Solar metallicity:
\HH\ becomes self-shielding at progressively larger N(H I) and smaller \HH\ fractions, but 
with nearly-fixed  $\AV = 0.15\pm0.03$ mag when comparing the Milky Way and Magellanic Clouds 
(Figure 5).  In Section 5 (see Figure 6) we showed that the onset of \HH\ self-shielding 
at progressively higher N(H I) and fixed \AV\ is a consequence of the smaller dust/gas ratio 
at smaller metallicity, and the consequent smaller grain surface area per unit hydrogen that 
is available for \HH\ formation.

In Section 6 (see Figure 7) we showed that higher N(HD)/N(\HH) and 
smaller N(\HH)/N(H) can be explained at higher density in systems of smaller 
metallicity, because the proton density increases with increasing number density,
opposite to the Milky Way case at Solar metallicity.  Moreover, in one case where
the J=1 level of HD is observed in a high-redshift system, the derived number
density n(H) $\approx 240 \pccc$ is relatively high.  However this system is 
not obviously of very low metallicity (Z/Z$_{\rm Solar} \la 0.36$) and high
N(HD)/N(\HH) ratios ranging from 80-200\% of the cosmic \DoverH\ are observed 
in high redshift systems over a wide range of metallicity ranging from 0.04 
to 2 times Solar (Table 1) for the 6 out of 8 high z systems with 
N(\HH) $> 8\times 10^{19}\pcc$ and smaller quoted column density errors.
Conversely, a low-redshift 
DLA system at z = 0.18 has N(HD) and N(\HH) values typical of Milky Way gas and 
a high \HH\ fraction, but at metallicity only 7\% Solar.

It seems odd that the high-redshift systems should share such high N(HD)/N(\HH)
values while having so little else in common in terms of the fractionation 
chemistry expected at their quoted metallicities.  For the high-z
systems it is tempting to abandon the fractionation scenario, with its implication
of a high H-ionization rate, in favor of an ad hoc scenario in which the 
molecules are in some tight knot where the conversion of both
hydrogen and deuterium to molecular form is  nearly complete.  But this begs
the question of why such similar and unusual conditions would exist over such
a wide range of metallicity.  

\subsection{\HH\ and the effect of dust extinction}

Motivated by effects like the sharp central upturns in HD/\HH\ observed in the upper 
panel of Figure 2, or as seen in Figure 1 at smaller \zetaH\ and high N(\HH),
we broke out the explicit extinction of \HH-dissociating Lyman and Werner Band
photons by dust.  As shown in Figures 2 and 3 and discussed in Sections 3 and 4,
the outer self-shielding layer in \HH\ is just that, triggered by non-linear 
effects inherent in the cross-section for \HH\ photo-absorption (Figure 3) with little
dependence on the existence of extinction by dust.  Only at very small n(H) in
a regime that is probably not thermally stable in two-phase ISM is the location 
of the onset of \HH-self shielding appreciably shifted by dust extinction.

By contrast, there is a somewhat wider range of number density where dust extinction 
has an appreciable effect on the \HH\ fraction inside the outermost \HH\ self-shielding
layer.  As shown in Figure 4, about one-third of the \HH\ in a cloud at n(H) = $32\pccc$
is directly attributable to dust extinction.

The effects of dust extinction on HD are somewhat greater owing to its
lesser degree of self-shielding.  As shown in Figure 2, HD/\HH\ ratios are
increased slightly even in the outer portions of our models, with much stronger
effects toward the center and even moreso at lower density.  Increases in
either the amount of \HH\ or the attenuation of HD-dissociating photons can
tip the HD over into a non-linear regime where its local abundance is strongly
dependent on its own self-shielding and somewhat less on the local fractionation
chemistry.

In discussing the dependence of \HH\ formation on metallicity, Figure 5 shows
how the location of the onset of strong \HH\ self-shielding shifts  monotonically
to higher N(H) with decreasing metallicity, in such a way as to keep constant the
inferred \AV\ representing the total column of grain surface area, see Sections
5 and 6.  This constancy of \AV\ could be interpreted as implying that the onset 
of \HH\ self-shielding is caused by the extinction due to dust, but that is not 
the case.  As shown in Figure 6, it is the lessening of the grain surface area 
available for \HH\ formation that causes the shift, while the dust extinction 
{\it per se}  increases the \HH\ fraction at yet-higher N(H), inside the \HH\ 
self-shielding layer.  As shown in Figure 6, producing \HH\ fractions above 
10\% requires even much higher N(H) in low-metallicity systems owing
to the diminished dust extinction cross section per H-atom.


\acknowledgments

  The National Radio Astronomy Observatory is currently  operated by Associated
  Universities, Inc. under a cooperative agreement with the National Science 
  Foundation.  The hospitality of the ITU-R and Hotel Bel Esperance in Geneva
  and the UKATC and the Apex City Hotel in Edinburgh are appreciated during the 
  writing of this manuscript.






\appendix

\section{Intra-cloud variation of the proton and electron densities}

Evaluating Eq 1 we find  n(p) $\approx 0.004\pccc$ for T = 75 K, using
the photodissociation rate of HD 
$\Gamma_{\rm HD} = \Gamma_{{\rm H}_2} = 4.3\times10^{-11}$\ps\ 
in free space \citep{DraBer96} and N(HD)/N(\HH) $\approx 3 \times10^{-6}$ as
shown in Figure 1.  This can be compared with the results shown in Figure 8,
where we have plotted the proton and electron densities just inside the
outer edge and at the center of the models whose results were depicted in
Figure 1.

The electron and proton densities behave very differently with 
respect to changes in density, largely because carbon remains ionized 
under all conditions, putting a floor on n(e) that does not exist for n(p).
At the outer edge, n(e) increases with density while n(p)
remains nearly fixed  owing to a balance between the volume ionization
n(H) \zetaH\ and the neutralization by small grains, whose density
and neutralization rate change in fixed proportion to n(H).
The electron density decreases toward the centers of all the models
because the ionization fraction  due to ionized hydrogen (ie, n(p))
decreases there.  The proton density decreases by nearly the same factor at 
all density while the decrease in the electron density at the center is small
at high density when ionization of hydrogen is weak.

The decline of n(p) toward the cloud center means that the chemistry has an 
innate tendency to segregate protons and \HH, shown in Figure 2, 
complicating the HD formation problem.   In Section 3.3 we noted that the 
distribution of HD is more uniform in radius at higher density, and the models
have higher \fH2\ at higher n(H)  but overall it is the proton density that
is decisive, and Figure 1 shows that lower density clearly wins the HD 
battle at Solar metallicity.  

\begin{figure}
 \includegraphics[height=10.5cm]{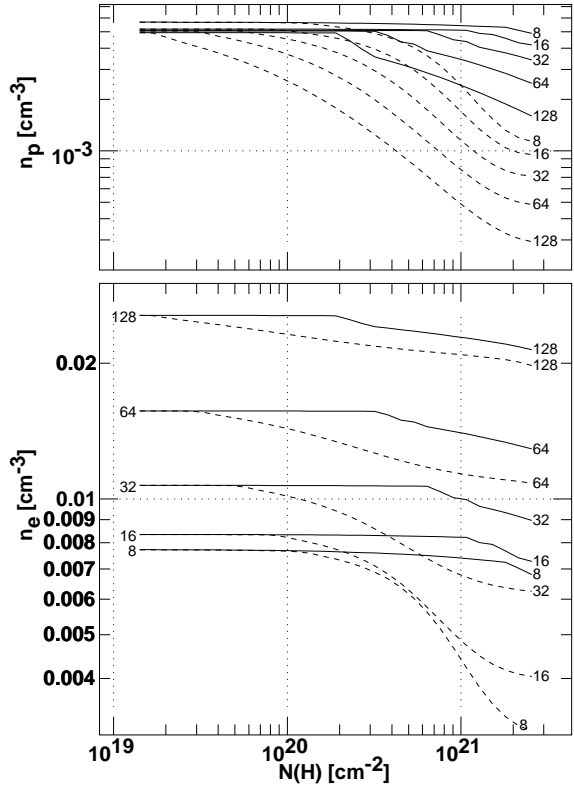}
  \caption[]{
Proton (upper) and electron densities for some of the models employed in 
Figure 1.  The solid lines are for locations just inside the outer
edge of the models; the dashed lines are at the center.  In all
cases the primary cosmic ray ionization rate per H-atom is
$\zetaH = 2\times 10^{-16}\ps$.  The electron and proton densities
move oppositely in response to changes in density at solar metallicity,
see Section 7.}
\end{figure}

\section{Scaling parameters for the \HH-formation problem}

Simply writing down the equation for the growth of \HH\ 

$$ dn(\HH)/dt = -n(\HH)\GH2 + \RH2 n(H) n(H I) \sqrt{\TK} \eqno(1) $$

and setting dn(\HH)/dt = 0 to achieve a formal solution (formal
because n(H) = n(H I) + 2n(\HH) and \GH2\ is functionally dependent
on N(\HH)) 

$$ n(\HH)/n(H I) = \RH2 \sqrt{\TK} n(H)/\GH2 $$
 
shows that $\Gamma_{\HH}/(\RH2 n(H) \sqrt{\TK}) $,  the n(\HH)/n(H I) 
ratio in free-space, is a dimensionless scaling parameter for this problem.  
It serves as the basis of the discussion of \cite{FedGla+79} who in terms of 
our quantities defined $1/\epsilon$ = n(H I)/2n(\HH) =
 $\Gamma_{\HH}/(2 \RH2 n(H) \sqrt{\TK})$
and showed that the data of \cite{SavDra+77} (see Figure 6 here)
could be reproduced with $\epsilon = 6\times 10^{-5}$ corresponding 
to n(H) $\approx 33 \pccc$ for \RH2 $\sqrt{\TK} = 3.9\times10^{-17}\pccc$
 \citep{GryBou+02} and $\GH2 = 4.3\times10^{-11}\ps$.   The formulation by 
\cite{FedGla+79} does not consider shielding due to dust 
and indeed, our models showed that dust shielding contributes modestly 
at that density at Solar metallicity (Section 4.2).   The formulation of 
\cite{FedGla+79} is numerically intensive but it provides a description of 
the width and location of  the self-shielding layer without additional assumptions 
because the equilibrium conditions are solved exactly, although not analytically.

The effects of dust extinction, shown in many instances in our models but
most important at small n(H), were
incorporated in the analytic formulation of Sternberg and his collaborators
\citep{Ste88,SteLeP+14} whose most recent description defines parameters
$\alpha = 2/\epsilon$ and G that is the mean \HH\ self-shielding factor
including the dust-extinction associated with N(\HH) (ie, not N(H)).  The
total dust column is considered to have two parts, called the H I and \HH\ dust,
whose columns are proportional to N(H I) and N(\HH). In this formulation the 
structure of the H I - \HH\ transition is determined by the product $\alpha$G.
The $\alpha$G $<< 1$ and $\alpha$G $>> 1$ limits are called the weak- and strong
field limits, respectively.  The domain where $\alpha$G $\approx 1$ marks 
the transition between H I and \HH-dominated regimes as indeed  seen from
the basic definition of $\alpha$ or $\epsilon$ as H I/\HH\ ratios;
one replaces the small free-space \HH/H I ratio by that in the strongly shielded
transition zone.   In effect, our models lie in the transition between 
two regimes, as understood in the context of the underlying physics of the
heating, cooling and ionization balance.

\cite{KMT08} explicitly introduced the total attenuation due to dust into
the equation of detailed balance (akin to eqn 1), resulting in the creation of 
a hybrid (chimerical?) parameter $\chi$ that is in effect a redefinition of 
$1/\epsilon$.  In $\chi$, the \HH\ photoabsorption cross section that determines 
\GH2\ is replaced by the cross-section for dust extinction.  Because the dust cross 
section is so much smaller, one deals with $\chi$-values of order 
unity in the H I - transition region as with Sternberg's formulation.
\cite{SteLeP+14} show that $\chi$ = $\alpha$G in the limit of low metallicity. 
 


\bibliographystyle{apj}

\end{document}